\documentclass[10pt,conference]{IEEEtran}
\IEEEoverridecommandlockouts

\usepackage{cite}
\usepackage{amsmath,amssymb,amsfonts}
\usepackage{algorithmic}
\usepackage{graphicx}
\usepackage{textcomp}
\usepackage{xcolor}
\usepackage[most]{tcolorbox}
\usepackage{todonotes}
\setuptodonotes{inline}
\def\BibTeX{{\rm B\kern-.05em{\sc i\kern-.025em b}\kern-.08em
    T\kern-.1667em\lower.7ex\hbox{E}\kern-.125emX}}
\begin{document}

\title{Revolutionizing Validation and Verification: Explainable Testing Methodologies for Intelligent Automotive Decision-Making Systems\\

}

\author{\IEEEauthorblockN{Halit Eris}
\IEEEauthorblockA{\textit{Technical University of Munich} \\
\textit{TUM School of Computation, Information and Technology}\\
Heilbronn, Germany \\
halit.eris@tum.de}
\and
\IEEEauthorblockN{Stefan Wagner}
\IEEEauthorblockA{\textit{Technical University of Munich} \\
\textit{TUM School of Computation, Information and Technology}\\
Heilbronn, Germany \\
stefan.wagner@tum.de}

}

\maketitle

\begin{abstract}
Autonomous Driving Systems (ADS) use complex decision-making (DM) models with multimodal sensory inputs, making rigorous validation and verification (V\&V) essential for safety and reliability. 
These models pose challenges in diagnosing failures, tracing anomalies, and maintaining transparency, with current manual testing methods being inefficient and labor-intensive. 
This vision paper presents a methodology that integrates explainability, transparency, and interpretability into V\&V processes. We propose refining V\&V requirements through literature reviews and stakeholder input, generating explainable test scenarios via large language models (LLMs), and enabling real-time validation in simulation environments. 
Our framework includes test oracle, explanation generation, and a test chatbot, with empirical studies planned to evaluate improvements in diagnostic efficiency and transparency. Our goal is to streamline V\&V, reduce resources, and build user trust in autonomous technologies.

\end{abstract}


\section{Introduction}
Autonomous driving systems (ADS) are transforming the automotive industry, increasing the demand for robust validation and verification (V\&V) methods to ensure safety, reliability, and compliance with user expectations \cite{1_10550712}. 
These systems use advanced decision-making (DM) algorithms, sometimes powered by artificial intelligence (AI) and sensor fusion from inputs like images, LIDAR data, speech, and text. 
As these technologies become integral to future transportation, rigorous testing is essential to guarantee their correct and reliable functioning under various conditions.

DM models for ADS process various sensory inputs, including images, videos, LIDAR point clouds, and multimodal data, to make complex decisions like navigation, localization, and motion planning \cite{2_10531702}. 
Their opacity complicates failure diagnosis and tracing root causes during V\&V. 
Ensuring effective V\&V is especially challenging and resource-intensive, as real-life tests or simulations often lack the transparency needed to observe DM processes \cite{3_10195149}.

Software engineers are vital in testing ADS through dynamic V\&V processes that include both real-world testing and simulations. 
Dynamic testing, as defined in the SWEBOK \cite{4_swebok}, involves executing the system under controlled conditions to observe its behavior and compare it to the expected outcomes. 
This process includes assessing the system's behaviors against user needs (validation) and predefined specifications and requirements (verification). 
However, managing unexpected or undesired behaviors during testing remains a major challenge. 
Currently, engineers document these anomalies using log files and technical reports, following ASPICE-compliant \cite{5_aspice} methodologies like the V-model. While systematic, these practices often lack transparency and fail to offer meaningful explanations to complex systems, making root cause analysis (RCA) labor-intensive and costly \cite{6_9153725}.

Imagine a scenario where an autonomous vehicle suddenly swerves off the road, potentially endangering its passengers. 
V\&V teams -- including software engineers, quality assurance teams, and AI developers -- often spend hours or even days analyzing extensive log files and documentation to diagnose the issue, unsure if the error lies in hardware, software, integration, or simulation tools. 
Explainable testing tools could provide a clear, interpretable explanation, such as: ``The decision was based on a misclassification of a LIDAR-detected object.'' Such insights would enable engineers to address problems quickly, saving time and reducing costs.

Current ADS test reports rely on low-level logs and post-mortem analyses, lacking structured insights into AI-driven DM. The absence of formalized explainability metrics hinders decision traceability \cite{15_sovrano2023objective}. 
Reports prioritize performance metrics over human-readable justifications, complicating failure analysis, increasing debugging costs, and delaying deployment. Integrating structured explanations -- RCA, decision rationales, and visual representations -- could improve efficiency and user trust \cite{16_preece2018asking}.

This vision paper addresses the challenges in ADS V\&V by proposing a methodology to make V\&V more efficient, transparent, and interpretable, thus enhancing user trust. 
Our approach integrates explainability using empirical research methods to develop a comprehensive V\&V tool as the ultimate contribution. 
We start by collecting requirements and empirical data and then design explainable scenarios for DM testing. 
The process includes creating datasets, a simulation environment, and a V\&V framework with test oracle, explanation and log generators, and an interactive test chatbot. 
Finally, we assess our methodology through statistical analysis and expert studies on ADS use cases. 
This vision outlines the following research contributions:

\begin{itemize}

\item \textbf{Explainable Testing Tool:} Develop a V\&V framework with a test oracle, explanation generator, and chatbot for clear, transparent insights.

\item \textbf{Scenario Generator:} Use LLMs to automatically create structured, explainable test scenarios.

\item \textbf{Interactive Explanations:} Implement a chatbot to provide easy-to-understand justifications for system behavior.

\item \textbf{Community Contribution:} Support the SE4ADS workshop by proposing a shared repository for datasets, test cases, and benchmarks.

\end{itemize}

\section{Related Work}

Thames et. al \cite{1_10550712} reviewed AI approaches for DM systems in safety-critical contexts, emphasizing that a key V\&V challenge is defining comprehensive requirements that fully capture system behavior. 
They suggest that future work should focus on addressing these unique challenges to develop systems that meet stringent safety needs.

Dennis and Fisher \cite{7_9094672} reviewed agent-based approaches for enhancing explainability in ADS, emphasizing methods like formal verification and self-simulation. 
They identified transparency in DM processes as a critical V\&V challenge and proposed integrating structured agent frameworks and runtime monitoring to improve failure diagnosis and safety. 
They also stressed the importance of ethical frameworks and continuous monitoring to align with societal norms. Their work includes a ``Why did you do that?'' feature, allowing users to question robotic actions using simplified Gwendolen semantics \cite{8_koeman2019did,10_techrep}, offering insights relevant to ADS.

Araluce et al. \cite{9_araluce2024leveraging} proposed a Transformer Encoder-based architecture to predict driver attention and explain vehicle decisions. Their approach enhances explainability in ADS by utilizing multiple annotated datasets that provide insights into driver behavior and accident causality. 
For instance, the Berkeley DeepDrive eXplanation (BDD-X) dataset contains textual ‘why’ explanations and heatmaps for DM analysis. 
The Detection of Traffic Anomaly (DoTa) dataset focuses on identifying anomalies with ‘what’ explanations and road annotations. The Comparison of Traffic Accidents (CTA) dataset captures causality in accidents with structured ‘why’ explanations. 
Similarly, the Honda Research Institute Driving (HDD) dataset describes driver actions, their causes, and attention patterns. 
Other datasets, such as BDD-A and BDD-OIA, offer first-person driver perspectives and explanations for complex maneuvers. 
By incorporating these datasets, the Transformer-based model learns to associate input sensory data with human-interpretable explanations, improving transparency in ADS DM.

While these studies offer valuable insights and directions regarding challenges, methodologies, and datasets, they do not prioritize software engineering or empirical research methodologies, and explainability has not been central to the testing of these systems. 
Despite existing research on V\&V challenges in ADS, there remains a critical gap in employing systematic software engineering methodologies and empirical research techniques to rigorously evaluate the impact of explainability on testing and diagnostics. 
With our proposed methodology, we aim to explore the impact of explainability on DM testing in the context of V\&V for ADS.

\section{Vision for Explainable Decision-Making System Testing in ADS Context}

We present a vision for an empirical research methodology to explore the impact of explainability in DM software testing and form an explainable testing tool. 
Our vision is built on key research questions (RQs) and four quality attributes: explainability, interpretability, transparency, and user trust. 

\subsection{Research Questions}

The following RQs guide the study and are central to its structure, ensuring that both the development of evaluation metrics and their validation through benchmarking and expert feedback are grounded in this empirical investigation.

\begin{tcolorbox}[colframe=black!75!white, colback=white, boxrule=0.5mm, sharp corners]
\textbf{RQ1}: What limitations and requirements do stakeholders face and verify in ADS testing?
\end{tcolorbox}

\begin{tcolorbox}[colframe=black!75!white, colback=white, boxrule=0.5mm, sharp corners]
\textbf{RQ2}: What factors affect the effectiveness of explainable V\&V methods in reducing diagnostic time and resources during real-life and simulation testing?
\end{tcolorbox}

\begin{tcolorbox}[colframe=black!75!white, colback=white, boxrule=0.5mm, sharp corners]
\textbf{RQ3}: How can state-of-art explainable mechanisms help testing teams understand the causes of unexpected system behaviors?
\end{tcolorbox}

\begin{tcolorbox}[colframe=black!75!white, colback=white, boxrule=0.5mm, sharp corners]
\textbf{RQ4}: How do explainable methods affect transparency and interpretability of V\&V, user trust, and acceptance of ADS among users and stakeholders?
\end{tcolorbox}

\begin{figure*}[htbp]
\centerline{\includegraphics[scale=0.135]{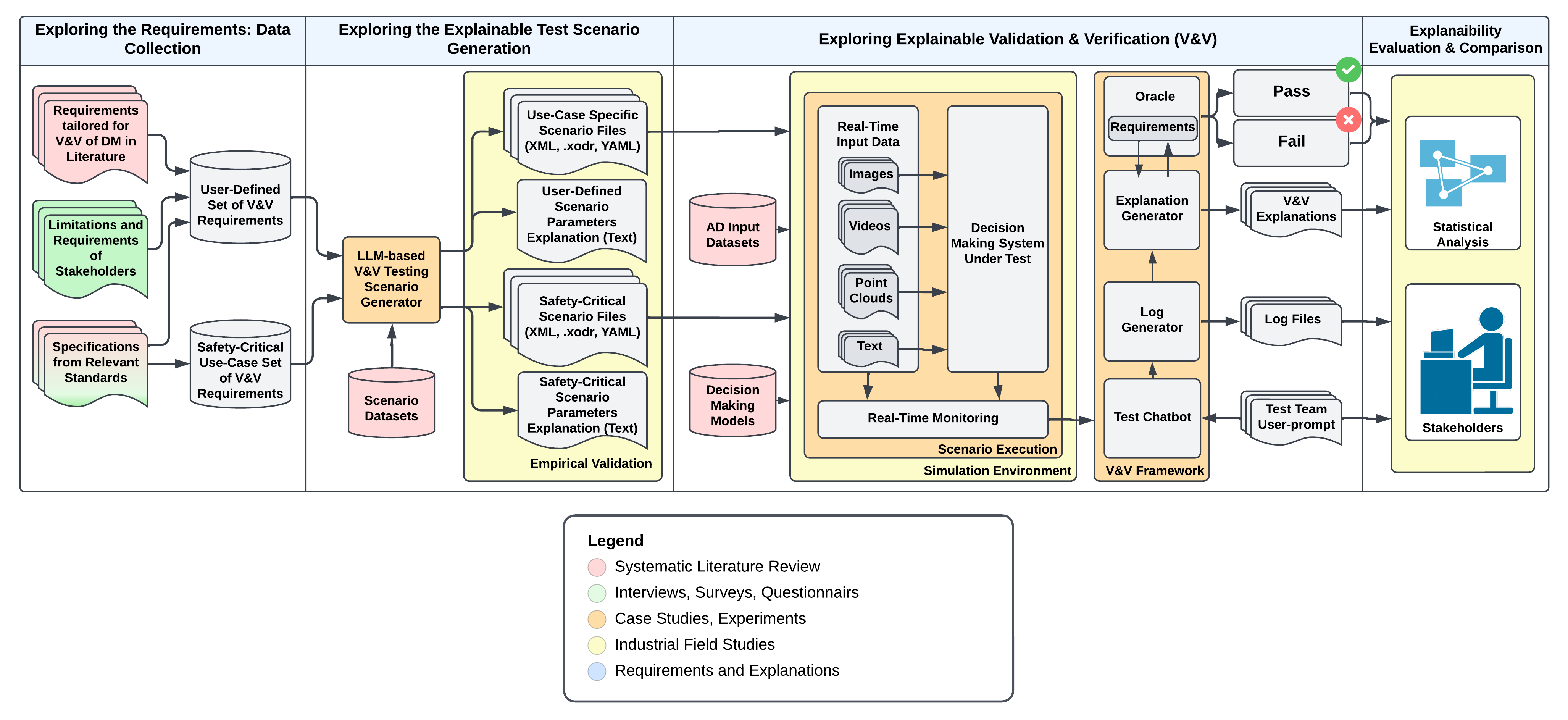}}
\caption{The proposed empirical methodology overview.}
\label{fig-1}
\end{figure*}

\subsection{Key Components of the Vision}

This subsection outlines the key components and their interrelations, forming our vision to achieve our objectives in ADS testing and development:

\begin{enumerate}
    \item \textbf{Explainability:} Explainability in ADS is the system's ability to clearly justify its decisions, helping testers and engineers diagnose and understand behavior. Using transparent information like sensor data, it offers justifications and simplifies outputs for non-experts, crucial for debugging and refining the system.

    \item \textbf{Transparency:} Providing access to the system’s algorithms, models, and data is crucial for regulatory compliance and effective debugging. Transparency supports explainability and interpretability, ensuring a full understanding of DM processes and fostering trust among regulators and engineers.

    \item \textbf{Interpretability:} Ensuring system outputs are easily understood by humans helps non-expert stakeholders, like drivers and passengers, grasp system actions, enabling ease of use and effective troubleshooting.
    
    \item \textbf{User Trust:} Trust relies on consistently applying explainability, transparency, and interpretability. Transparency reassures technical stakeholders through verifiability, while explainability and interpretability boost user confidence by making system behavior understandable and predictable, encouraging adoption and safe use.
    
\end{enumerate}

\subsection{Proposed Methodology Overview}

Our proposed methodology consists of four phases, as illustrated in Fig.\ref{fig-1}. Each phase uses specific empirical methods aimed at testing and validating DM systems for ADS and contributes to building a comprehensive V\&V tool. The legend in the figure highlights the empirical methods applied at each step. The phases are described in detail below, with an expected outcome for each in an example case of testing DM for Automated Emergency Braking (AEB):

\subsubsection{Phase 1: Exploring the Requirements -- Data Collection}

We begin by identifying and collecting V\&V requirements tailored to DM systems in ADS to establish a solid foundation for our methodology addressing the RQ1 and RQ2. 
A systematic literature review (SLR) \cite{11_guide} will gather existing requirements and industry standards from academic and professional sources. 
In parallel, we will conduct quantitative and qualitative studies, such as interviews, surveys, and questionnaires, to supplement the findings of the literature. 

This phase will produce a comprehensive set of user-defined and safety-critical requirements. In this vision paper we share an example requirement: ``The AEB system shall be tested on a controlled track with both standard driving and high-speed inter-urban conditions to ensure it activates effectively when approaching a stationary vehicle, particularly at speeds of 30-50 km/h, where driver warnings are less effective.''

\subsubsection{Phase 2: Exploring Explainable Test Scenario Generation}

From the text-based requirements in Phase 1, we will create an explainable test scenario generator. 
Leveraging LLM-based models like ChatGPT, Gemini, or LLaMA, the generator will automatically produce safety-critical, use-case-specific scenarios in structured formats (XML, xodr, YAML) with parameter explanations \cite{12_tian2025lmmenhancedsafetycriticalscenariogeneration, 13_Zhao_2024}. 

\begin{tcolorbox}[colframe=black!75!white, colback=white, boxrule=0.5mm, sharp corners]

\textbf{Scenario Explanation:}

\textbf{[Vehicle-A]} travels in the right lane of an [Urban Road] at \textbf{[40 km/h]}, positioned \textbf{[100 meters]} from \textbf{[Vehicle-B]} with a \textbf{2-meter} safe stopping distance. The scenario tests \textbf{[Automated Emergency Braking]} for DM Model \textbf{[Model-1]}, using Camera and LIDAR, with monitoring of key parameters: \textbf{[List-Parameters]}.
\end{tcolorbox}

Using Everything as Code (EaC) \cite{14_eac} will automate documentation and streamline integration into development pipelines. 
We will validate this generator through case studies and experiments, curate scenario datasets via SLR, and perform empirical validation in industrial field studies to ensure practical effectiveness.

\subsubsection{Phase 3: Exploring Explainable Validation \& Verification (V\&V)}

In this phase, we will use pre-existing AD input datasets and models from the SLR to test DM systems. 
A simulation environment built from industrial field studies will allow real-time monitoring of inputs like images, videos, and point clouds. 
We will validate scenarios and analyze system behavior through case studies and experiments, refining monitoring parameters as the research progresses. 
A test chatbot will explain results, providing insights and justifications based on prompts, with the structure of these prompts optimized during the study.

\begin{tcolorbox}[colframe=black!75!white, colback=white, boxrule=0.5mm, sharp corners]

\textbf{System Explanation:} Emergency braking was delayed due to a LIDAR misclassification error prioritized by the sensor fusion algorithm, causing an unsafe stopping distance.

\textbf{Model Behavior Explanation:} To reduce false positives in urban settings, the AEB model favored LIDAR data over the camera, compromising safety when misclassification occurred.

\textbf{Monitored Parameters:} LIDAR Confidence: 80\% (Misclassified), Camera Confidence: 90\% (Correct), Distance to Vehicle B: 1m (Actual) vs. 2m (Expected), Braking Delay: 300ms (due to sensor fusion misalignment)

\end{tcolorbox}

\subsubsection{Phase 4: Explainability Evaluation \& Comparison}

The final phase evaluates our V\&V methods' explainability through expert feedback from industry professionals, addressing RQ3 and RQ4, conducting industrial field studies to assess key performance and explainability metrics, using statistical analysis to refine these metrics based on research insights.

\section{Community Infrastructure Contribution}

To support the SE4ADS workshop's goal of an ADS-specific repository, we propose collaborating to standardize resources on explainability methodologies. 
Engaging participants, we aim to curate datasets, test cases, and benchmarks, fostering a shared infrastructure for reproducible research. 
This repository will facilitate meaningful technique comparisons and advance innovation in explainable V\&V for ADS.

\section{Conclusion}
We propose a new methodology to improve the V\&V processes of ADS DM systems. Our approach integrates explainability, transparency, and interpretability, addressing current challenges faced in testing and debugging. 
This methodology spans requirement gathering, scenario generation, validation, and explainability evaluation, aiming to enhance efficiency and user trust. 
We believe this work can serve as a foundational step toward more reliable and comprehensible ADS testing practices, ultimately benefiting software engineers and the automotive industry.

\section*{Acknowledgment}
This work was partially supported by the Ministry of Science,
Research and Arts of the Federal State of Baden-Württemberg
in the project SdMobi5 –- TESSOF
within the Innovations Campus Mobilität der Zukunft and by the German Federal Ministry of Education and Research in the project AutoDevSafeOps (01IS22087R).

\bibliographystyle{IEEEtran}
\bibliography{IEEEabrv,bibmain}

\begin{thebibliography}{10}
\providecommand{\url}[1]{#1}
\csname url@samestyle\endcsname
\providecommand{\newblock}{\relax}
\providecommand{\bibinfo}[2]{#2}
\providecommand{\BIBentrySTDinterwordspacing}{\spaceskip=0pt\relax}
\providecommand{\BIBentryALTinterwordstretchfactor}{4}
\providecommand{\BIBentryALTinterwordspacing}{\spaceskip=\fontdimen2\font plus
\BIBentryALTinterwordstretchfactor\fontdimen3\font minus \fontdimen4\font\relax}
\providecommand{\BIBforeignlanguage}[2]{{%
\expandafter\ifx\csname l@#1\endcsname\relax
\typeout{** WARNING: IEEEtran.bst: No hyphenation pattern has been}%
\typeout{** loaded for the language `#1'. Using the pattern for}%
\typeout{** the default language instead.}%
\else
\language=\csname l@#1\endcsname
\fi
#2}}
\providecommand{\BIBdecl}{\relax}
\BIBdecl

\bibitem{1_10550712}
C.~Thames and Y.~Sun, ``A survey of artificial intelligence approaches to safety and mission-critical systems,'' in \emph{2024 Integrated Communications, Navigation and Surveillance Conference (ICNS)}, 2024, pp. 1--12.

\bibitem{2_10531702}
X.~Zhou, M.~Liu, E.~Yurtsever, B.~L. Zagar, W.~Zimmer, H.~Cao, and A.~C. Knoll, ``Vision language models in autonomous driving: A survey and outlook,'' \emph{IEEE Transactions on Intelligent Vehicles}, pp. 1--20, 2024.

\bibitem{3_10195149}
M.~Geisslinger, R.~Trauth, G.~Kaljavesi, and M.~Lienkamp, ``Maximum acceptable risk as criterion for decision-making in autonomous vehicle trajectory planning,'' \emph{IEEE Open Journal of Intelligent Transportation Systems}, vol.~4, pp. 570--579, 2023.

\bibitem{4_swebok}
H.~Washizaki, \emph{Guide to the Software Engineering Body of Knowledge (SWEBOK Guide), Version 4.0}.\hskip 1em plus 0.5em minus 0.4em\relax IEEE Computer Society, 2024.

\bibitem{5_aspice}
\BIBentryALTinterwordspacing
\emph{Automotive SPICE® 4.0}, VDA QMC, 2023, release 4.0. [Online]. Available: \url{https://vda-qmc.de/en/automotive-spice/automotive-spice-veroeffentlichungen/. [Accessed: Jan. 28, 2025]}
\BIBentrySTDinterwordspacing

\bibitem{6_9153725}
Y.~Wang, J.~J. Yang, and N.~M. Mbiye, ``An automotive ehps software reliability and testing,'' in \emph{2020 Annual Reliability and Maintainability Symposium (RAMS)}, 2020, pp. 1--6.

\bibitem{15_sovrano2023objective}
F.~Sovrano and F.~Vitali, ``An objective metric for explainable ai: How and why to estimate the degree of explainability,'' \emph{Knowledge-Based Systems}, vol. 278, p. 110866, 2023.

\bibitem{16_preece2018asking}
A.~Preece, ``Asking ‘why’in ai: Explainability of intelligent systems--perspectives and challenges,'' \emph{Intelligent Systems in Accounting, Finance and Management}, vol.~25, no.~2, pp. 63--72, 2018.

\bibitem{7_9094672}
L.~A. Dennis and M.~Fisher, ``Verifiable self-aware agent-based autonomous systems,'' \emph{Proceedings of the IEEE}, vol. 108, no.~7, pp. 1011--1026, 2020.

\bibitem{8_koeman2019did}
V.~J. Koeman, L.~A. Dennis, M.~Webster, M.~Fisher, and K.~Hindriks, ``The “why did you do that?” button: Answering why-questions for end users of robotic systems,'' in \emph{International Workshop on Engineering Multi-Agent Systems}.\hskip 1em plus 0.5em minus 0.4em\relax Springer, 2019, pp. 152--172.

\bibitem{10_techrep}
L.~A. Dennis, ``Gwendolen semantics: 2017,'' University of Liv- erpool, Department of Computer Science, Tech. Rep. ULCS-17-001, 2017.

\bibitem{9_araluce2024leveraging}
J.~Araluce, L.~M. Bergasa, M.~Oca{\~n}a, {\'A}.~Llamazares, and E.~L{\'o}pez-Guill{\'e}n, ``Leveraging driver attention for an end-to-end explainable decision-making from frontal images,'' \emph{IEEE Transactions on Intelligent Transportation Systems}, 2024.

\bibitem{11_guide}
D.~Budgen and P.~Brereton, ``Performing systematic literature reviews in software engineering,'' in \emph{Proceedings of the 28th International Conference on Software Engineering}, ser. ICSE '06.\hskip 1em plus 0.5em minus 0.4em\relax New York, NY, USA: Association for Computing Machinery, 2006, p. 1051–1052.

\bibitem{12_tian2025lmmenhancedsafetycriticalscenariogeneration}
\BIBentryALTinterwordspacing
H.~Tian, X.~Han, Y.~Zhou, G.~Wu, A.~Guo, M.~Cheng, S.~Li, J.~Wei, and T.~Zhang, ``Lmm-enhanced safety-critical scenario generation for autonomous driving system testing from non-accident traffic videos,'' 2025. [Online]. Available: \url{https://arxiv.org/abs/2406.10857}
\BIBentrySTDinterwordspacing

\bibitem{13_Zhao_2024}
\BIBentryALTinterwordspacing
Y.~Zhao, W.~Xiao, T.~Mihalj, J.~Hu, and A.~Eichberger, ``Chat2scenario: Scenario extraction from dataset through utilization of large language model,'' in \emph{2024 IEEE Intelligent Vehicles Symposium (IV)}.\hskip 1em plus 0.5em minus 0.4em\relax IEEE, Jun. 2024, p. 559–566. [Online]. Available: \url{http://dx.doi.org/10.1109/IV55156.2024.10588843}
\BIBentrySTDinterwordspacing

\bibitem{14_eac}
\emph{Everything as code - DevOps Guidance}, AWS Documentation, 2024, release 1.0.

\end{thebibliography}

\end{document}